\documentclass[twocolumn,  
 showpacs,  
 preprintnumbers,  
 aps,  
 prd,  
 a4paper,  
 superscriptaddress,  
 nofootinbib,  
 tightenlines,  
 floats,floatfix  
 ]{revtex4}

\usepackage{amsmath}
\usepackage{amsfonts}
\usepackage{graphicx}

\begin{document}

\title{Two-Field Quintom Models in the $w-w'$ Plane}

\bigskip

\author{Zong-Kuan Guo}
\email{guozk@itp.ac.cn}
\affiliation{Institute of Theoretical Physics, Chinese Academy of
     Sciences, P.O. Box 2735, Beijing 100080, China}

\author{Yun-Song Piao}
\email{yspiao@gucas.ac.cn}
\affiliation{College of Physical Sciences, Graduate School of Chinese
     Academy of Sciences, YuQuan Road 19A, Beijing 100049, China}

\author{Xinmin Zhang}
\affiliation{Institute of High Energy Physics, Chinese Academy of
     Sciences, P.O. Box 918-4, Beijing 100039, China}

\author{Yuan-Zhong Zhang}
\affiliation{CCAST (World Lab.), P.O. Box 8730, Beijing 100080, China}
\affiliation{Institute of Theoretical Physics, Chinese Academy of
     Sciences, P.O. Box 2735, Beijing 100080, China}

\begin{abstract}
 The $w-w'$ plane, defined by the equation of state parameter for
 the dark energy and its derivative with respect to the logarithm
 of the scale factor, is useful to the study of classifying the
 dynamical dark energy models. In this note, we examine
 the evolving behavior of the two-field quintom models with $w$
 crossing the $w=-1$ barrier in the $w-w'$ plane. We find that
 these models can be divided into two categories,
 type A quintom in which $w$ changes from $>-1$ to $<-1$
 and type B quintom in which $w$ changes from $<-1$ to $>-1$ as the
 universe expands.
\end{abstract}


\maketitle


Recent observations of type Ia supernovae suggest that the
expansion of the universe is accelerating and that two-thirds
of the total energy density exists in a dark energy component
with negative pressure~\cite{rie98}.
In addition, measurements of the cosmic microwave
background~\cite{spe03} and the galaxy power spectrum~\cite{teg04}
also indicate the existence of the dark energy.
The simplest candidate for the dark energy is a cosmological
constant $\Lambda$, which has pressure
$P_\Lambda=-\rho_\Lambda$. Specifically, a reliable model
should explain why the present amount of the dark
energy is so small compared with the fundamental scale
(fine-tuning problem) and why it is comparable with the
critical density today (coincidence problem).
The cosmological constant suffers from both these problems.
One possible approach to constructing a viable model for
dark energy is to associate it with a slowly evolving and
spatially homogeneous scalar field $\phi$, called
``quintessence''~\cite{rat88,zla99,fer98}. Such a model
for a broad class of potentials can give the energy density
converging to its present value for a wide set of initial
conditions in the past and possess tracker behavior
(see, e.g.,~\cite{sah00} for a review).

Quintessence models describe the dark energy with a time-varying
equation of state parameter, $w$, the ratio of its pressure to
energy density and $w>-1$. Recently, such models have been
extended to phantom dark energy with $w<-1$~\cite{cal02} (see for
example~\cite{one04} and references therein). The physical sources
for phantom fields with strongly negative pressure may be looked
for in string theory~\cite{mer01} and supergravity~\cite{nil84}.
Such fields may arise from quantum effects on a locally de Sitter
background~\cite{one02}. They may also be present in modified
gravity theories, such as higher order theories~\cite{pol88} and
scalar-tensor theories~\cite{tor02}. Coupled quintessence with
dark matter may also lead to $w<-1$~\cite{car05}. Phantom field
with a negative kinetic term may be a simplest implementing, in
which the weak energy condition is violated. It has been shown
that such models possess the attractor behavior similar to
quintessence models~\cite{guo04}.

If $w < -1$ in an expanding universe, the energy density of the
dark energy increases with time, which leads to unwanted future
singularity called ``big rip''~\cite{cal03}. Thus from this point
of view the transition from $w > -1$ to $w < -1$ or vice versa
would be desirable for the history of the universe. On the other
hand, the analysis on the properties of dark energy from the
recent observations mildly favors models with $w$ crossing $-1$ in
the near past. From the theoretical viewpoint, it is necessary to
explore possibilities for dark energy with $w$ crossing $-1$.
However, neither quintessence nor phantom can fulfill this
transition. The similar conclusion has also been obtained for the
k-essence models~\cite{vik04}. Quintom models easily provide a way
to realize this transition~\cite{fen05,guo05}. The quintom fields
may be associated with some higher derivatives terms~\cite{li05}
derived from fundamental theories, for instance due to the quantum
corrections or the non-local physics in the string
theory~\cite{sim90}. They may also arise from a slowly decaying
D3-brane in a local effective approximation~\cite{are05}.
Interestingly, the quintom models differ from the quintessence or
phantom in evolution and the determination of the fate of
universe~\cite{fen04}. There exist lots of interests in the
literature presently in building of quintom-like
models~\cite{wei04}, such as hessence models~\cite{wei05} and
brane models~\cite{cof05}.

Recently, Caldwell and Linder examined the evolving behavior
of quintessence models of dark energy in the $w-w'$ phase plane,
where $w'$ is the time derivative of $w$ with respect to the
logarithm of the scale factor $a$, and showed that these models
occupy the thawing and freezing regions in the phase plane~\cite{cal05}.
More recently, these results were extended to a more general
class of quintessence models with a monotonic potential~\cite{sch05}
and phantom dark energy~\cite{chi05}.
In this note we extend these studies with single-field quintessence or
phantom models
to two-field quintom models. Our results show that there exist two
types of quintom models according to the evolving behavior around $w=-1$.
Moreover, we plot the trajectories numerically for the two types
in the $w-w'$ plane.


Let us consider the following model which contains a negative-kinetic
scalar field $\phi$ (phantom) and a normal scalar field $\psi$
(quintessence):
\begin{eqnarray}
S &=& \int d^4x\sqrt{-g}\Bigg[\frac{R}{2\kappa ^2}
 -\frac{1}{2}g^{\mu \nu}\partial _\mu \phi \partial _\nu \phi
 +\frac{1}{2}g^{\mu \nu}\partial _\mu \psi \partial _\nu \psi
 \nonumber\\
&& \hspace{20mm} +V(\phi,\psi)+\mathcal{L_{\textrm{m}}}\Bigg],
\end{eqnarray}
where $\kappa^2 \equiv 8\pi G_N$ the gravitational coupling,
$V(\phi,\psi)$ the scalar potential and
$\mathcal{L_{\textrm{m}}}$ the Lagrangian density of matter fields.
In a flat FRW cosmology the evolutions of the fields are governed by
\begin{eqnarray}
\label{em1}
&& \ddot{\phi} + 3H\dot{\phi} - V_{,\phi}=0\,, \\
&& \ddot{\psi} + 3H\dot{\psi} + V_{,\psi}=0\,,
\label{em2}
\end{eqnarray}
where $V_{,\phi}=\partial V/\partial \phi$ and
$V_{,\psi}=\partial V/\partial \psi$. In what follows
we assume that there is no direct coupling between the phantom field
and the normal scalar field, i.e.,
$V(\phi,\psi)=V_{\phi}(\phi)+V_{\psi}(\psi)$.
Then the effective equation of state $w$ is given by
\begin{eqnarray}
\label{ES2}
w&=& \frac{-\dot{\phi}^2 + \dot{\psi}^2 - 2V}
 {-\dot{\phi}^2 + \dot{\psi}^2 + 2V}\,, \\
&=& \frac{\Omega_{\phi}w_{\phi}+\Omega_{\psi}w_{\psi}}
   {\Omega_{\mathrm{DE}}}\,,
\label{ES}
\end{eqnarray}
where $\Omega_{\rm DE}=\Omega_{\phi}+\Omega_{\psi}$ is the
density parameter of dark energy,
$w_{\phi}=(\dot{\phi}^2+2V_{\phi})/(\dot{\phi}^2-2V_{\phi})$
and $w_{\psi}=(\dot{\psi}^2-2V_{\psi})/(\dot{\psi}^2+2V_{\psi})$.
For a model with a normal scalar field, the equation of state
$w \ge -1$. The toy model of a phantom energy component with a negative
kinetic term possesses an equation of state $w < -1$. In our model,
Eq.~(\ref{ES}) implies $w \ge -1$ when the quintessence component $\psi$
dominates the density of the universe and $w<-1$ when
the phantom component $\phi$ dominates.
By using the equations of motion (\ref{em1}) and (\ref{em2}),
the derivative of (\ref{ES2}) with respect to $\ln a$ can be
rewritten as~\cite{zla99}
\begin{equation}
w' = 3(1-w)^2\frac{\dot{\phi}^2-\dot{\psi}}{2V}
 - (1-w)\frac{1}{V}(V_{,\phi}\frac{\dot{\phi}}{H}
 + V_{,\psi}\frac{\dot{\psi}}{H}).
\end{equation}
By using the following relations
\begin{eqnarray}
&&\frac{\dot{\phi}^2-\dot{\psi}}{2V}=-\frac{1+w}{1-w}\,,\quad
\frac{\kappa \dot{\phi}}{H}=\pm \sqrt{-3(1+w_\phi)\Omega_\phi}
\,, \nonumber \\
&&\frac{\kappa \dot{\psi}}{H}=\pm \sqrt{3(1+w_\psi)\Omega_\psi}\,,
\end{eqnarray}
the expression for $w'$ becomes
\begin{eqnarray}
\label{weq}
w'&=&-3(1-w^2) + (1-w)\frac{1}{\kappa V}
 \Bigg[\pm V_{,\phi} \sqrt{-3(1+w_\phi)\Omega_\phi}
 \nonumber\\
&& \hspace{20mm}\pm V_{,\psi} \sqrt{3(1+w_\psi)\Omega_\psi}
 \Bigg]\,.
\end{eqnarray}
For a phantom field $\phi$ climbing up its potential, the $\pm$ sign
before $V_{,\phi}$ depends on whether $V_{,\phi}<0$ or
$V_{,\phi}>0$, respectively.
For a quintessence field $\psi$ rolling down its potential,
the $\pm$ sign before $V_{,\psi}$ corresponds to
$V_{,\psi}>0$ or $V_{,\psi}<0$, respectively.

According to the evolving behavior of $w$ around $-1$,
the two-field quintom models of dynamical dark energy
are classified into the following two types:
type A quintom characterized by $w$ from $w>-1$ to $w<-1$
and type B quintom characterized by $w$ from $w<-1$ to $w>-1$.

{\bf Type A quintom models:}

\begin{figure*}[t]
\begin{center}
\includegraphics[angle=-90,width=16cm]{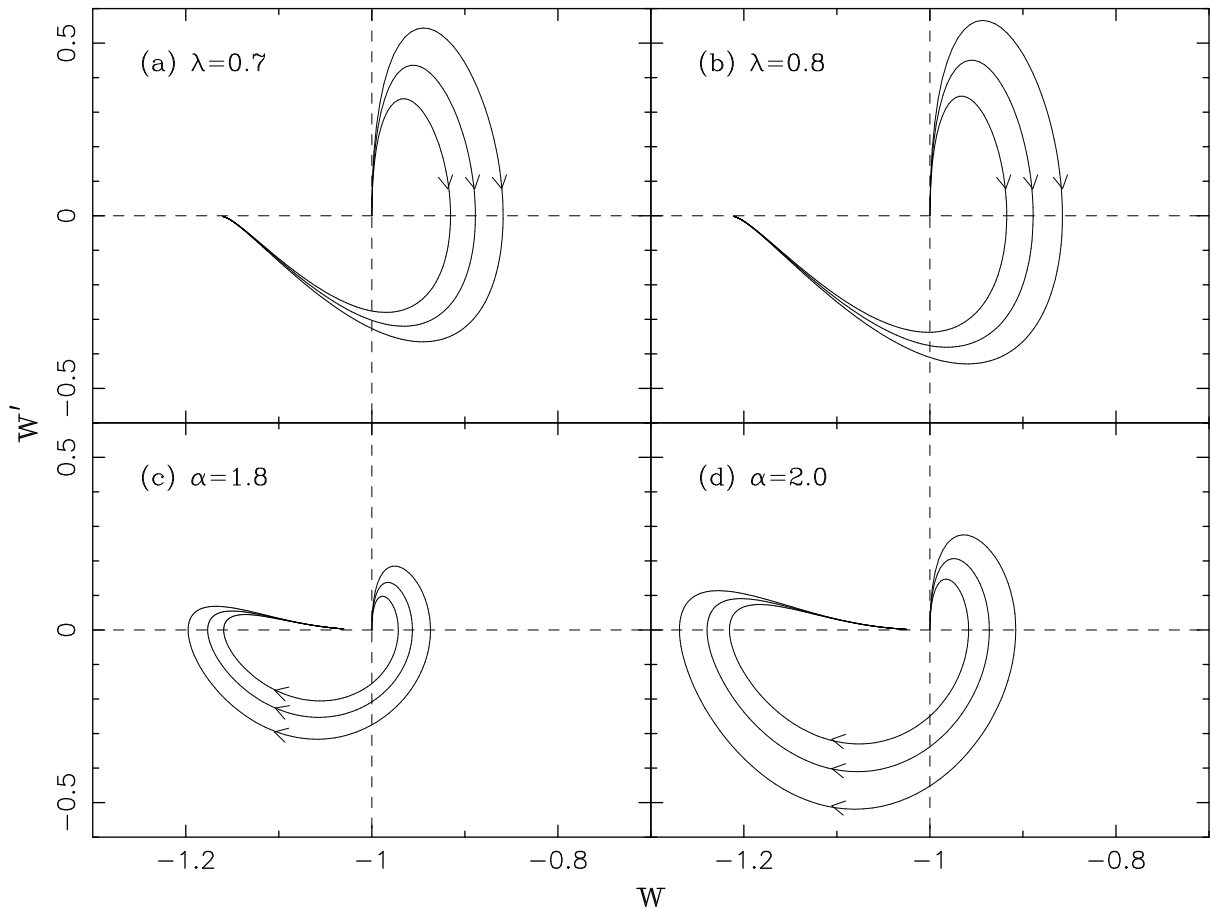}
\caption{The evolution of the type A quintom models of dynamical dark
 energy in the $w-w'$ phase plane. The tracks in panels (a), (b),
 (c) and (d) correspond to the quintom models with the potentials
 $V(\phi,\psi)=V_{\phi 0} e^{-\lambda_{\phi} \phi}
 +V_{\psi 0} e^{-\lambda_{\psi} \psi}$
 ($\lambda_{\phi}=0.7$, $0.8$)
 and $V(\phi,\psi)=V_{\phi 0}\phi^{\alpha}+V_{\psi 0}\psi^{\alpha}$
 ($\alpha=1.8$, $2.0$), respectively.}
\label{fig:f1}
\end{center}
\end{figure*}

In such models, the equation of state changes from $w>-1$ to $w<-1$,
i.e., the universe evolves from a quintessence-dominated phase to
a phantom-dominated phase.
Therefore, the properties of the late-time attractor solution are
determined by the phantom potential $V_{\phi}(\phi)$.
The cosmological evolution of the Type A quintom model with
\begin{equation}
V(\phi,\psi)=V_{\phi 0} \,e^{-\lambda_{\phi}\kappa \phi}
 + V_{\psi 0} \,e^{-\lambda_{\psi} \kappa \psi}
\end{equation}
was investigated in detail in Ref.~\cite{guo05}.
When the phantom component is dominated at late times,
Eq.~(\ref{weq}) reduces to
\begin{equation}
w'=(1-w)\left[-3(1+w)-\lambda_{\phi}\sqrt{-3(1+w)} \right],
\end{equation}
which has three critical points $w=1$, $w=-1$
and $w=-1-\lambda_{\phi}^2/3$. It is easily shown that the scaling
solution $w=-1-\lambda_{\phi}^2/3$ is the stable attractor of this
type of models, i.e., the ratio of kinetic to potential energy of
the phantom field becomes a constant.
The top panels of Fig.~\ref{fig:f1} shows the evolving
behavior of the models in the $w-w'$ phase plane.
We shall next consider a general case in which the effective
equation of state $w$ tends to $-1$ at late times.
In this case, the ratio of kinetic to potential energy tends
to zero~\cite{guo04}.
The features of the behavior are virtually independent of
the precise shape of the quintessence potential since the
contribution of the quintessence component becomes negligible
at late times compared to the phantom component.
For example we consider a positive power-law potential,
which have been previously investigated in
Refs.~\cite{sam04,guo04}.
For the following potential
\begin{equation}
V(\phi,\psi)=V_{\phi 0}\phi^{\alpha}
 +V_{\psi 0}\psi^{\alpha},
\end{equation}
in the $\phi \to \infty$ limit, we have $w'=-3(1-w)(1+w)$.
The critical point with $w=-1$ is the late-time attractor,
i.e., the quintom field becomes ultimately frozen, as shown
in the bottom panels of Fig.~\ref{fig:f1}.

{\bf Type B quintom models:}

\begin{figure*}[t]
\begin{center}
\includegraphics[angle=-90,width=16cm]{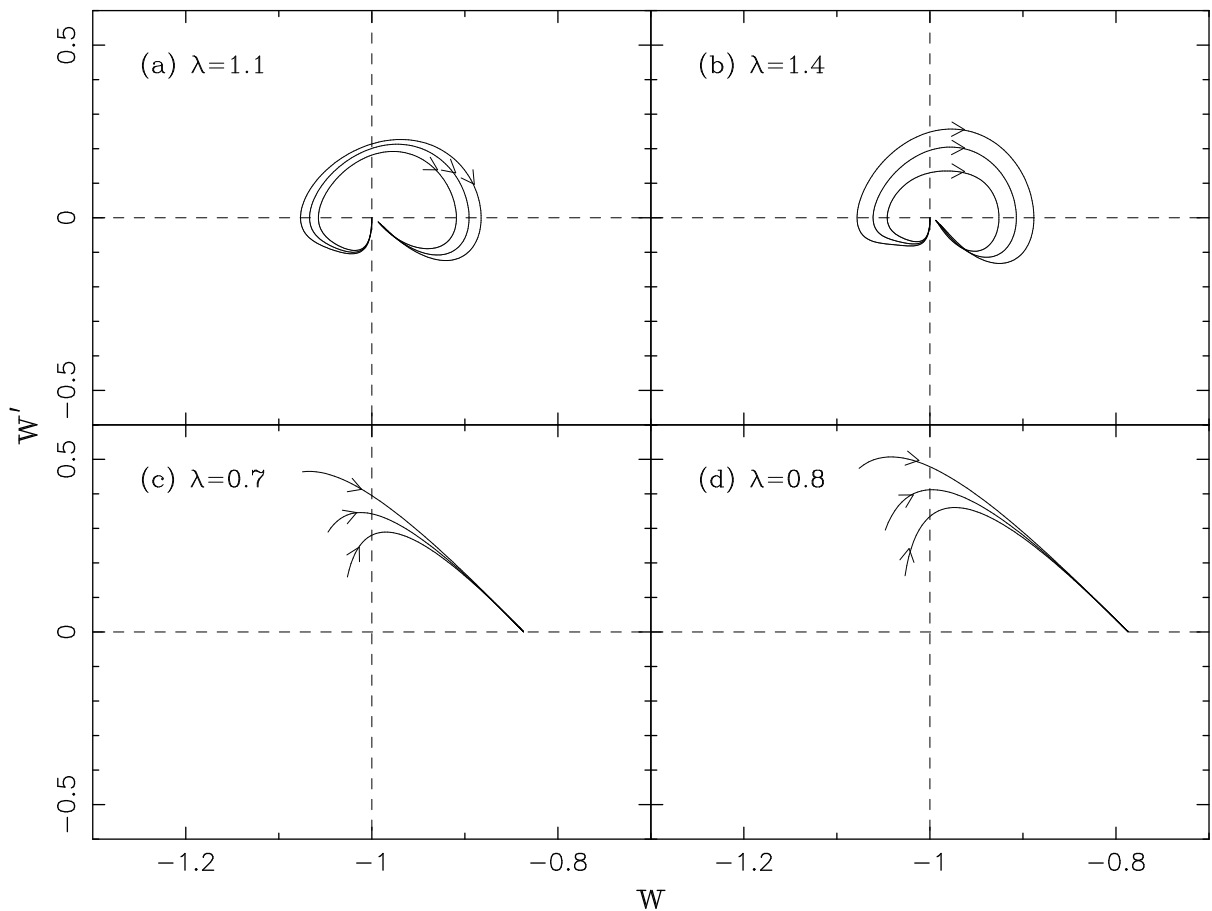}
\caption{The evolution of the type B quintom models of dynamical dark
 energy in the ($w, w'$) phase plane. The tracks in panels (a), (b),
 (c) and (d) correspond to the quintom models with the potentials
 $V(\phi,\psi)=V_{\phi 0} e^{-\lambda_{\phi} \phi^2}
 +V_{\psi 0} e^{-\lambda_{\psi} \psi^2}$ ($\lambda_{\psi}=1.1$, $1.4$)
 and  $V(\phi,\psi)=V_{0} e^{-\lambda \psi}$ ($\lambda=0.7$, $0.8$),
 respectively.}
\label{fig:f2}
\end{center}
\end{figure*}

In such models, the equation of state changes from $w<-1$ to
$w>-1$. In principle, for the two-field system consisting of
quintessence and phantom, if the phantom initially dominates the
universe, it will still dominate up to future, and thus the
universe can not exit the phantom phase ($w<-1$) forever.
We take a simple potential
\begin{equation}
V(\phi,\psi)=V_{\phi 0} e^{-\lambda_{\phi}
\phi^2} + V_{\psi 0} e^{-\lambda_{\psi} \psi^2}
\end{equation}
for a illustration.
The relevant figure is plotted in the bottom panels of
Fig.~\ref{fig:f2}.
In general the phantom field will climb up its potential, which
makes its energy density increasing constantly during its
evolution. Thus in order to implement the change of equation of
state from $w<-1$ to $w>-1$, a naive choice is taking the
potential of phantom field zero, for example, considering the
following potential
\begin{equation}
V(\phi,\psi)=V_{\psi 0}e^{-\lambda_{\psi} \psi},
\end{equation}
whose evolving behavior is plotted
in the top panels of Fig.~\ref{fig:f2}. In this case,
the initial kinetic energy of phantom field is not zero, which
naturally leads
to the state equation $w<-1$ of quintom system. But with the
expansion of the universe, the kinetic energy of phantom field
will become negligible gradually while the energy of
quintessence field will begin to dominate the universe. Thus we
have $w>-1$. However, it should be noted that here the energy of
phantom field is negative, but the total energy of quintom system
is still positive. We do not want to deeply discuss the relevant
physics with this case, and we here only emphasize that to obtain
such a transition from $w<-1$ to $w>-1$ in two-field system
(quintessence+phantom), the above condition seems to be necessarily
satisfied.

In conclusion, it is well-known that the two-field quintom models
give a simplest realization to the dynamical dark energy with the
equation of state parameter crossing the $w=-1$ barrier,
thus it is very interesting to explore the full evolving behavior
of various two-field quintom models.
In this note, we have examined the evolving behavior of the
two-field quintom models in the $w-w'$ plane.
We find that these
models can generally be divided into two categories,
type A quintom models in which $w$ changes from $>-1$ to $<-1$ and
type B quintom models in which $w$ changes from $<-1$ to $>-1$,
which has not been analyzed in detail before.
Compared to the latter, the former is easily constructed
since the energy density of the phantom field increases
as the universe expands and ultimately dominates the universe.
The latter requires the phantom field with a flat potential or
a potential with a maximum.

\section*{Acknowledgements}
This project was in part supported by National
Basic Research Program of China under Grant No. 2003CB716300 and
by NNSFC under Grant No. 90403032.


\end{document}